%% file: eprint_arxiv.tex
\documentclass[12pt]{article}
\usepackage{graphicx}
\usepackage{amsmath}
\usepackage{subfig}

\newcommand\pubnumber{SNSN-???-??}
\newcommand\pubdate{\today}

\def\LBNL{Lawrence Berkeley National Laboratory,\\
1 Cyclotron Road, Berkeley, CA, USA}

\def\Title#1{\begin{center} {\Large #1 } \end{center}}
\def\Author#1{\begin{center}{ \sc #1} \end{center}}
\def\Address#1{\begin{center}{ \it #1} \end{center}}

\newcommand\pubblock{\rightline{\begin{tabular}{l} \pubnumber\\
         \pubdate  \end{tabular}}}
\newenvironment{Abstract}{\begin{quotation}  }{\end{quotation}}
\newenvironment{Presented}{\begin{quotation} \begin{center} 
             PRESENTED AT\end{center}\bigskip 
      \begin{center}\begin{large}}{\end{large}\end{center} \end{quotation}}

\input econfmacros.tex

\begin{document}
\begin{titlepage}
\pubblock

\vfill
\Title{Latest Updates from the AlCap Experiment}
\vfill
\Author{Andrew Edmonds\\on behalf of the AlCap Collaboration}
\Address{\LBNL}
\vfill
\begin{Abstract}
The AlCap experiment is a joint venture between the COMET and Mu2e collaborations that will measure the rate and spectrum of particles emitted after nuclear muon capture on aluminium. Both collaborations will search for the charged lepton flavour violating process of neutrinoless muon-to-electron conversion by stopping muons in an aluminium target. Knowledge of other particles emitted during this process is important. The AlCap charged particle emission data was collected at the Paul Scherrer Institut in Switzerland over two runs in 2013 and 2015. In this talk, the experiment will be described and the current status will be presented.
\end{Abstract}
\vfill
\begin{Presented}
Conference on the Intersections of Particle and Nuclear Physics\\
Palm Springs, USA,  May 29--June 3, 2018
\end{Presented}
\vfill
\end{titlepage}
\def\thefootnote{\fnsymbol{footnote}}
\setcounter{footnote}{0}

\section{Introduction}
The next generation of muon-to-electron conversion searches will search for the charged lepton flavour violating process of neutrinoless $\mu-e$ conversion with a single event sensitivity of $\sim10^{-17}$. The rate of this process is normalised to the rate of the nuclear muon capture process, i.e.:

\begin{center}
  $R_{\mu\rightarrow e} = \dfrac{\Gamma[\mu^{-} + N(Z, A) \rightarrow e^{-} + N(Z, A)]}{\Gamma[\mu^{-} + N(Z, A) \rightarrow \nu_{e} + N(Z-1, A)]}$.
\end{center}

\begin{figure}[htb]
\centering
\subfloat[][Diagram of COMET Phase-II.\label{fig:comet}]{\includegraphics[width=0.5\textwidth]{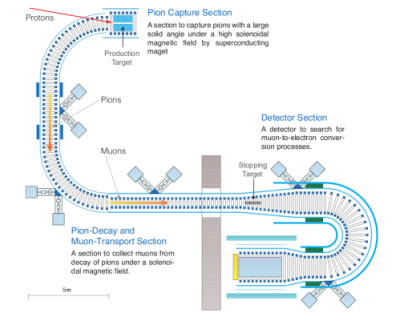}}
\subfloat[][Diagram of Mu2e.\label{fig:mu2e}]{\includegraphics[width=0.5\textwidth]{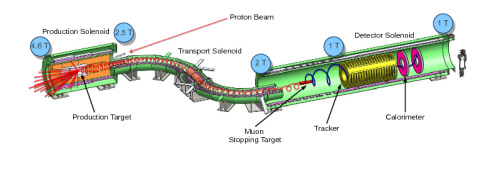}}
\caption{Diagrams of the next generation of $\mu-e$ conversion experiments.}
\end{figure}

The signal for $\mu-e$ conversion is a monoenergetic electron of 105 MeV and two experiments that will search for this process are COMET~\cite{comet} at J-PARC in Japan (Fig.~\ref{fig:comet}) and Mu2e~\cite{mu2e} at Fermilab in the USA (Fig.~\ref{fig:mu2e}). These experiments have slightly different experimental designs but their underlying method is the same: produce a large flux of muons ($O(10^{10})$ per second), stop them in an aluminium stopping target and measure the momentum of any emitted electrons to find the 105 MeV/c signal.

Along with any signal electrons, other particles will be emitted from the stopping target. X-rays are produced when a muon is stopped by an atom and falls into the $1s$ orbit; background electrons are produced when a muon decays normally to an electron and two neutrinos; and gamma rays, heavy charged particles and neutrons are produced when a muon is captured by the nucleus and the excited nuclear state relaxes back to the ground state.

All these particles are important to Mu2e/COMET. Any heavy charged particle with a momentum within the detector acceptance (a proton of 5 MeV has a momentum of 100 MeV/c) will deposit large amounts of energy, which could damage the detectors. Neutrons will travel out of the experimental area, pass through the cosmic ray vetos and fake a cosmic ray trigger which would increase the deadtime of the experiments. Finally, and more usefully, both the X-rays that are emitted from the initial stop and the gamma rays that are emitted from the capture process can be used to determine the denominator of $R_{\mu\rightarrow e}$. 

\begin{figure}[htb]
\centering
\includegraphics[width=\textwidth]{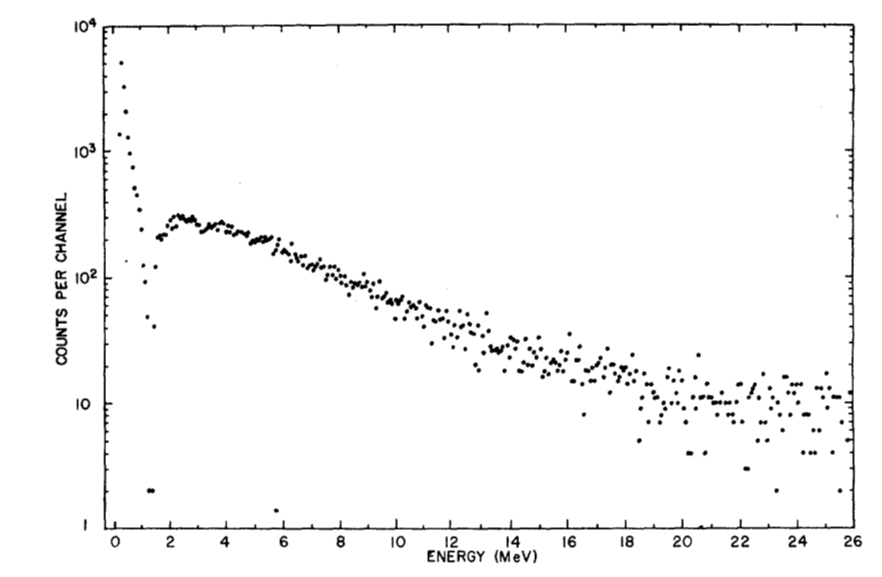}
\caption{The charged particle emission spectrum for particles emitted after nuclear muon capture on silicon~\cite{sobottka}.}
\label{fig:sobottka}
\end{figure}

Currently, data exists on charged particle emission from nuclear muon capture on aluminium at high energies ($E>40$ MeV)~\cite{wyttenbach}, which is too high for COMET/Mu2e, and at low energies is only known for muonic silicon~\cite{sobottka} (shown in Fig.~\ref{fig:sobottka}). In both these measurements, there is no information on the composition of the emitted charged particles, which will be useful for Mu2e/COMET to know. For neutrons, the rate is known~\cite{Macdonald:1965zz} but not the spectrum; and for photons, the X-ray and gamma ray energies are known but there are large uncertainties on the intensities of the gamma rays~\cite{Measday:2007zz}. For these reasons the AlCap experiment~\cite{alcap} was performed, with members from both the COMET and Mu2e collaborations, to determine the rate and spectrum of individual charged particles, neutrons and photons after nuclear muon capture.

\section{The AlCap Experiment}
The AlCap experiment collected data in three runs at the Paul Scherrer Institut (PSI) in Switzerland. The first run in 2013 focussed on charged particles and photons with preliminary work done on the neutron measurement. The second run was performed in summer 2015 and focussed on the neutron and photon measurement. Finally the third run, in winter 2015, focussed again on charged particles and photons and collected data from aluminium (50 $\mu$m, 100 $\mu$m); silicon (52 $\mu$m and 1500 $\mu$m), to cross-check our analyses against literature; and titanium (50 $\mu$m), which is an alternate stopping target material for COMET/Mu2e. This proceeding will focus on the charged particle analysis of the 52 $\mu$m silicon dataset from the 2015 run.

\begin{figure}[htb]
\centering
\includegraphics[width=\textwidth]{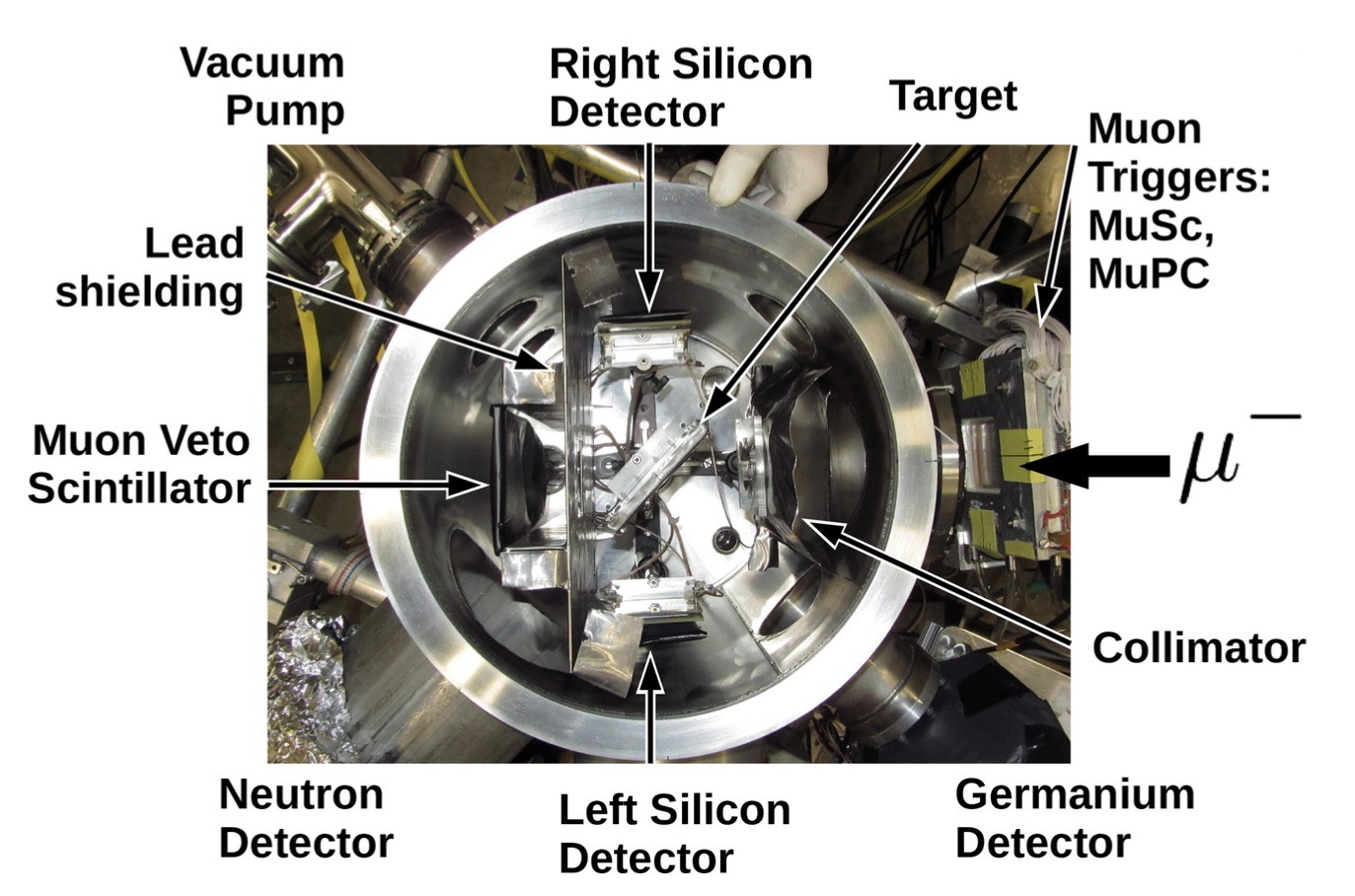}
\caption{Photo of the AlCap experimental setup in 2013.}
\label{fig:alcap}
\end{figure}

The experimental setup for the charged particle data is shown in Fig.~\ref{fig:alcap}. A stopping target is placed at the centre of a vacuum chamber with two detector packages on either side. In addition, there is a muon counter at the entrance to the chamber to record the entering muon, and a germanium detector outside the chamber, pointing towards the target, to measure the stopping X-rays for normalisation. 

\begin{figure}[htb]
\centering
\includegraphics[width=\textwidth]{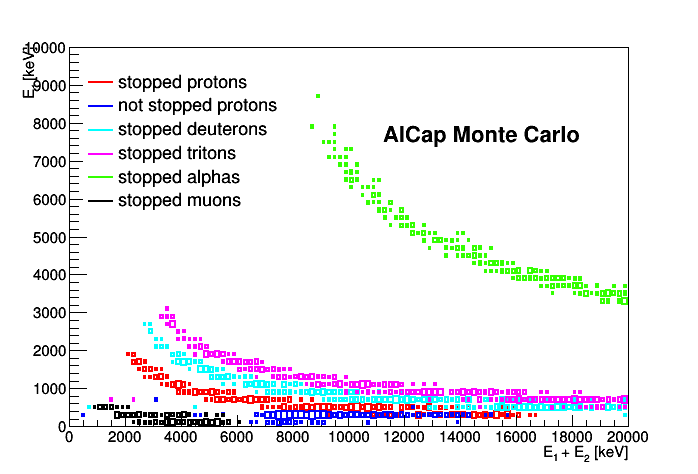}
\caption{Plot of $E_{1}$ vs $E_{1}+E_{2}$ from Monte Carlo simulations of different particle types being emitted from a stopping target and measured in the silicon detector packages.}
\label{fig:pid-demo}
\end{figure}

For the 2015 dataset, the detector packages on either side of the target consisted of a thin silicon layer (50 $\mu$m) and two thick silicon detectors (1500 $\mu$m). With this arrangement, particle identification can be performed by plotting the energy deposited in the thin layer against the sum of the energy deposited in the thin and thick layers. The plot in Fig.~\ref{fig:pid-demo} from our Monte Carlo simulations demonstrates this for a two layer analysis.

\section{Charged Particle Analysis}
\subsection{Silicon Dataset}
\subsubsection{Protons}
\begin{figure}[htb]
\centering
\subfloat[][Plot of $E_{1}$ vs $E_{1}+E_{2}$.\label{fig:si16b-evde}]{\includegraphics[width=0.5\textwidth]{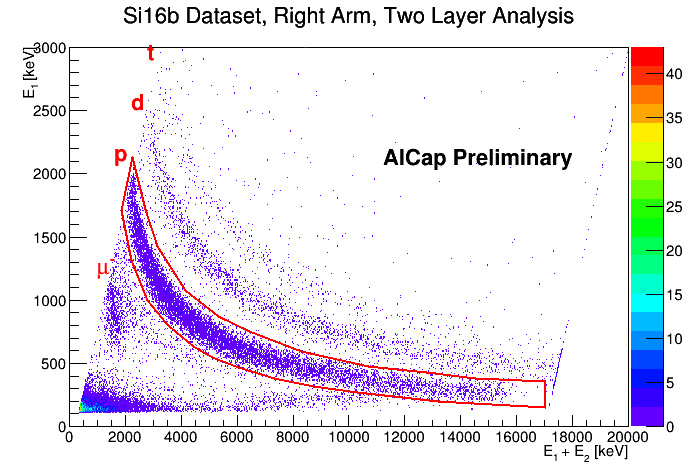}}
\subfloat[][Plot of proton arrival times.\label{fig:si16b-proton-time}]{\includegraphics[width=0.5\textwidth]{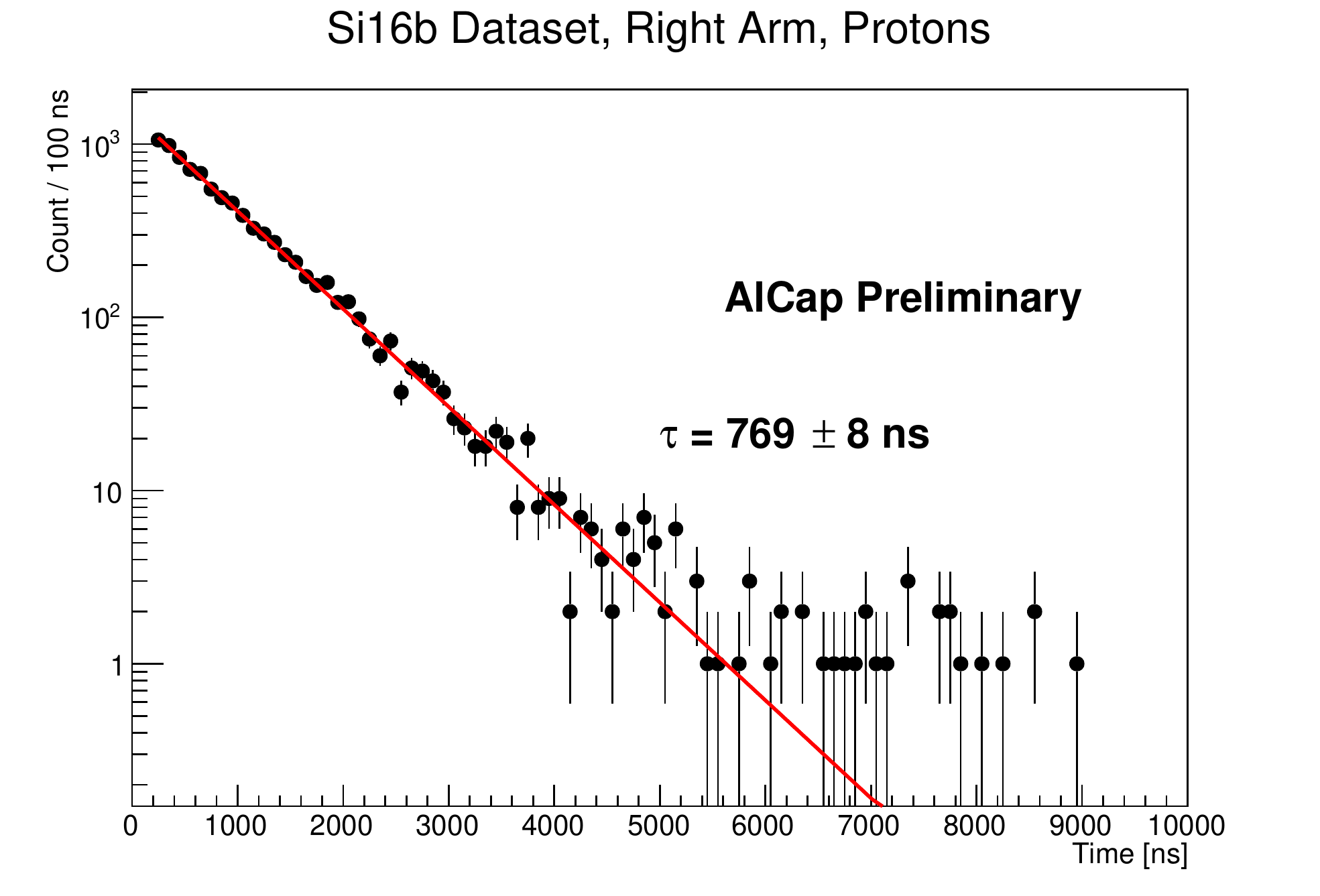}}
\caption{Plots from the silicon charged particle analysis. (Left): Plot of $E_{1}$ vs $E_{1}+E_{2}$ and the proton band extraction cut (red). (Right): Plot of arrival times (relative to the incoming muon) of protons extracted from (left) showing a decay time consistent with the lifetime of muonic silicon (767 ns).}
\end{figure}

Fig.~\ref{fig:si16b-evde} shows the $E_{1}$ vs $E_{1}+E_{2}$ plot for this dataset and the individual particle bands are clear. The proton band is extracted with simple cuts and a plot of the arrival time of the extracted protons (Fig.~\ref{fig:si16b-proton-time}) fits to an exponential decay with a lifetime consistent with literature (767 ns~\cite{suzuki}). Work is ongoing to determine the efficiency and purity of the proton band cut.

\begin{figure}[htb]
\centering
\includegraphics[width=1.0\textwidth]{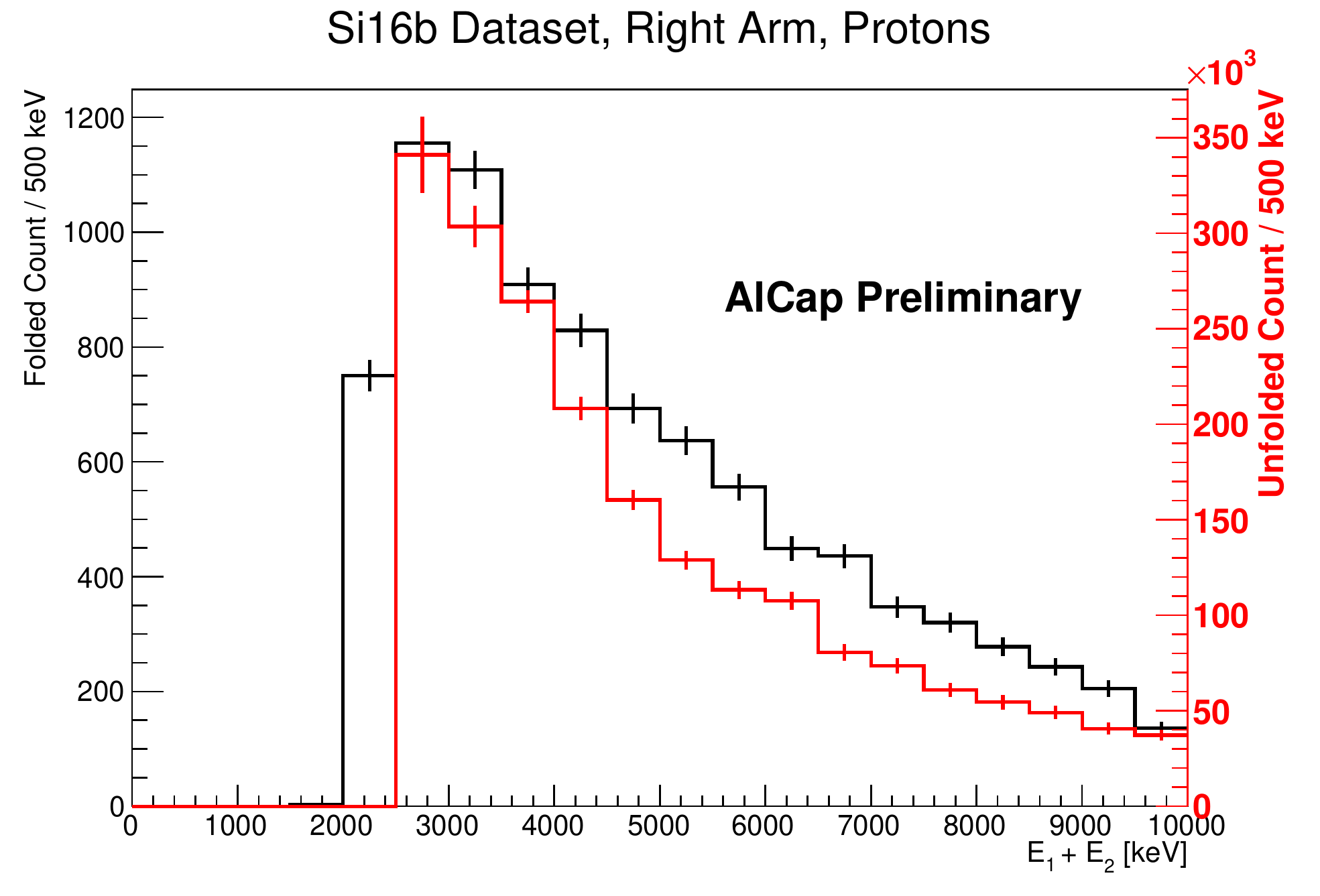}
\caption{Proton energy spectrum before unfolding (black, left axis) and after unfolding (red, right axis) of protons emitted after nuclear muon capture on silicon.}
\label{fig:si16b-proton-spectra}
\end{figure}

A Bayesian unfolding method is used to account for the geometric efficiency and energy loss of the protons as they leave the stopping target. These effects are extracted from a Monte Carlo simulation, where a muon beam, tuned to data, is fired at the stopping target to produce a stopping position distribution. Protons are generated uniformly in energy along this position distribution and a response matrix, which maps the generated energy against the recorded energy, is created. This response matrix is used as an input to the unfolding program~\cite{roounfold} along with the $x$-projection of the extracted proton band to get the unfolded energy spectrum. Fig.~\ref{fig:si16b-proton-spectra} shows the folded (black) and unfolded (red) spectra of protons in this dataset. Since the unfolding relies on the muon beam simulation, which has uncertainties, work is being carried out to investigate how the unfolded spectrum changes as the stopping distribution changes.

\subsubsection{Normalisation}
\begin{figure}[htb]
\centering
\includegraphics[width=1.0\textwidth]{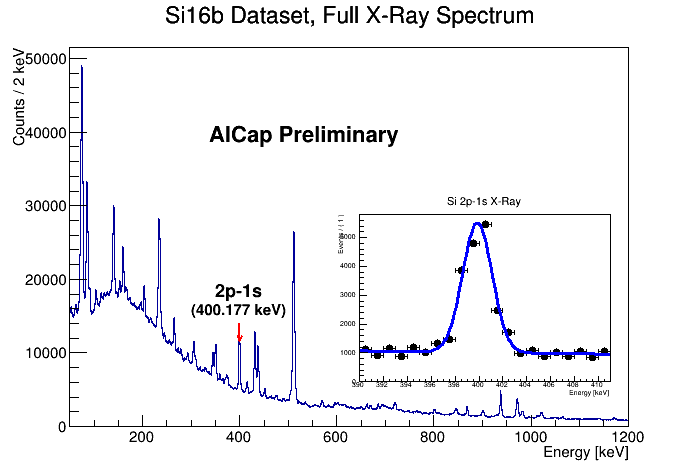}
\caption{X-ray spectrum for the silicon dataset and fit to the silicon $2p-1s$ peak (inset).}
\label{fig:si16b-full-xray}
\end{figure}

The rate of proton emission is normalised to the number of captured muons, which cannot be counted directly in this experiment. Instead, the number of stopped muons and the fraction of stopped muons that are captured by the nucleus (0.658 for silicon~\cite{Measday:2007zz}) are used. The number of stopped muons can be counted by fitting to the $2p-1s$ peak of the X-ray spectrum (Fig.~\ref{fig:si16b-full-xray}).

\section{Conclusion}
The AlCap experiment was performed by members of both the Mu2e and COMET collaborations to measure the rate and spectrum of charged particles, neutrons and photons after nuclear muon capture.

This proceeding presented the current status of the charged particle analysis on silicon. The aluminium and titanium datasets are being analysed in parallel and results from all materials should be completed soon.

\end{document}

%% file: econfmacros.tex
\def\beq{\begin{equation}}
\def\eeq#1{\label{#1}\end{equation}}
\def\eeqn{\end{equation}}

\def\beqa{\begin{eqnarray}}
\def\eeqa#1{\label{#1}\end{eqnarray}}
\def\eeqan{\end{eqnarray}}

\let\bar=\overbar

\def\Dslash{\not{\hbox{\kern-4pt $D$}}}
\def\dslash{\not{\hbox{\kern-2pt $\del$}}}

\def\msb{{\bar{\ssstyle M \kern -1pt S}}}

